%% file: spherical_mirror.tex
\documentclass[12pt]{article}
\usepackage{graphicx}
\usepackage{cite}
\usepackage{color}
\newcommand{\bm}[1]{\mbox{\boldmath $#1$}}
\newcommand{\fnd}[2]{\frac{\textstyle #1}{\textstyle #2}}
\newcommand{\fndrs}[4]{\fnd{\raisebox{#1}{$#2$}}{\raisebox{#3}{$#4$}}}
\newcommand{\xrm}[1]{{\textstyle \mbox{\rm #1}}}

\newcommand{\abs}[1]{\left| #1\right|}
\newcommand{\x}[1]{{\textstyle #1}}

\begin{document}
\title{Images in Christmas Balls}
\author{
Eef van Beveren\\
{\normalsize\it Centro de F\'{\i}sica Te\'{o}rica}\\
{\normalsize\it Departamento de F\'{\i}sica, Universidade de Coimbra}\\
{\normalsize\it P-3004-516 Coimbra, Portugal}\\
{\bf\footnotesize http://cft.fis.uc.pt/eef}\\ [10pt]
\and
Frieder Kleefeld and George Rupp\\
{\normalsize\it Centro de F\'{\i}sica das Interac\c{c}\~{o}es Fundamentais}\\
{\normalsize\it Instituto Superior T\'{e}cnico, Edif\'{\i}cio Ci\^{e}ncia}\\
{\normalsize\it P-1049-001 Lisboa Codex, Portugal}\\
{\bf\footnotesize kleefeld@cfif.ist.utl.pt}\\
{\bf\footnotesize george@ist.utl.pt}\\ [10pt]
{\small PACS number(s): 01.40.Ej, 01.40.Fk, 42.15.Dp, 42.15.Fr}
}
\date{\today}
\maketitle

\begin{abstract}
We describe light-reflection properties of spherically curved mirrors, like
balls in the Christmas tree. In particular, we study the position of the image
which is formed somewhere beyond the surface of a spherical mirror,
when an eye observes the image of a pointlike light source.
The considered problem, originally posed by Abu Ali Hasan Ibn al-Haitham ---
alias Alhazen --- more than a millennium ago, turned out to have
the now well known analytic solution of a biquadratic equation, being still
of great relevance, e.g.\ for the aberration-free construction of telescopes.
We do not attempt to perform an exhaustive survey of the rich historical
and engineering literature on the subject, but develop a simple pedagogical
approach to the issue, which we believe to be of continuing interest in view
of its maltreating in many high-school textbooks.
\end{abstract}
\clearpage

\section{Introduction}

The basic property of geometrical optics for reflection is the equality
of the angles of incidence and reflection, given by the equation
\begin{equation}
t\; =\; i
\;\;\; .
\label{teqi}
\end{equation}
Here, we will demonstrate that equality (\ref{teqi}) is not respected
when analyzing curved mirrors, in particular spherical mirrors,
in the educational programs for high-school and university students
\cite{spmi}.

One of the corner stones of geometrical optics at the high-shool level
is the concept of ideal lens,
which reduces the lens to an optical plane,
principal and auxiliary axes,
and a focal plane with principal and auxiliary foci.
The rules for image formation are clear:
incident light rays parallel to an auxiliary axis are refracted in the lens
such that the refracted light ray passes through the corresponding
auxiliary focus in the case of a positive ideal lens, or seem to originate
from the corresponding auxiliary focus for a negative ideal lens.
No further rules are needed to construct images for any ideal lens or
set of ideal lenses.
This concept is not only useful, as it approximates well real lenses
applied in optical instruments, but also a practical application of
Euclidean geometry.
\clearpage

\begin{figure}[htbp]
\begin{center}
\begin{tabular}{|p{240pt}|p{180pt}|}
\hline & \\
\centerline{\scalebox{0.5}{\includegraphics{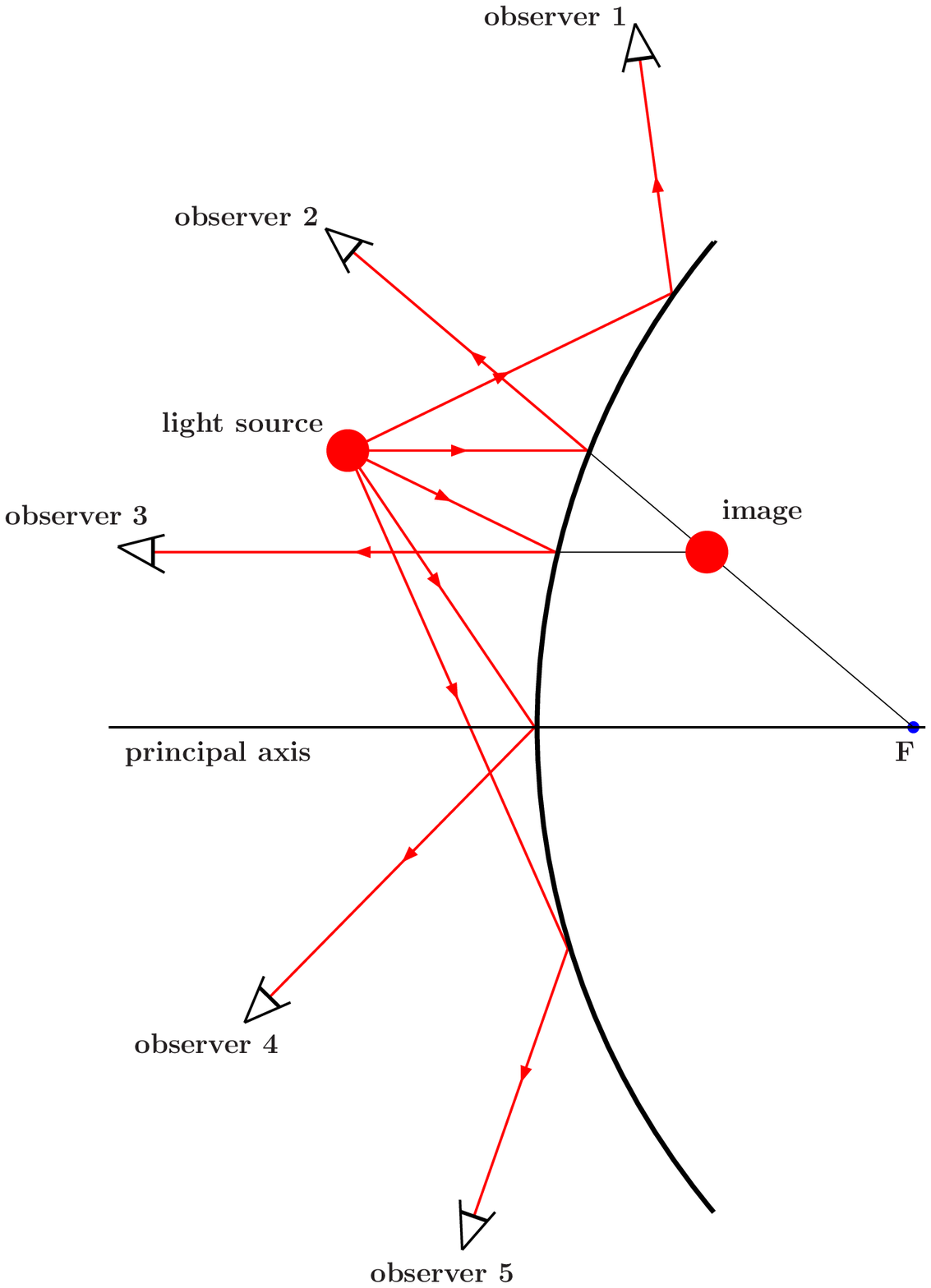}}} &
\centerline{\scalebox{0.5}{\includegraphics{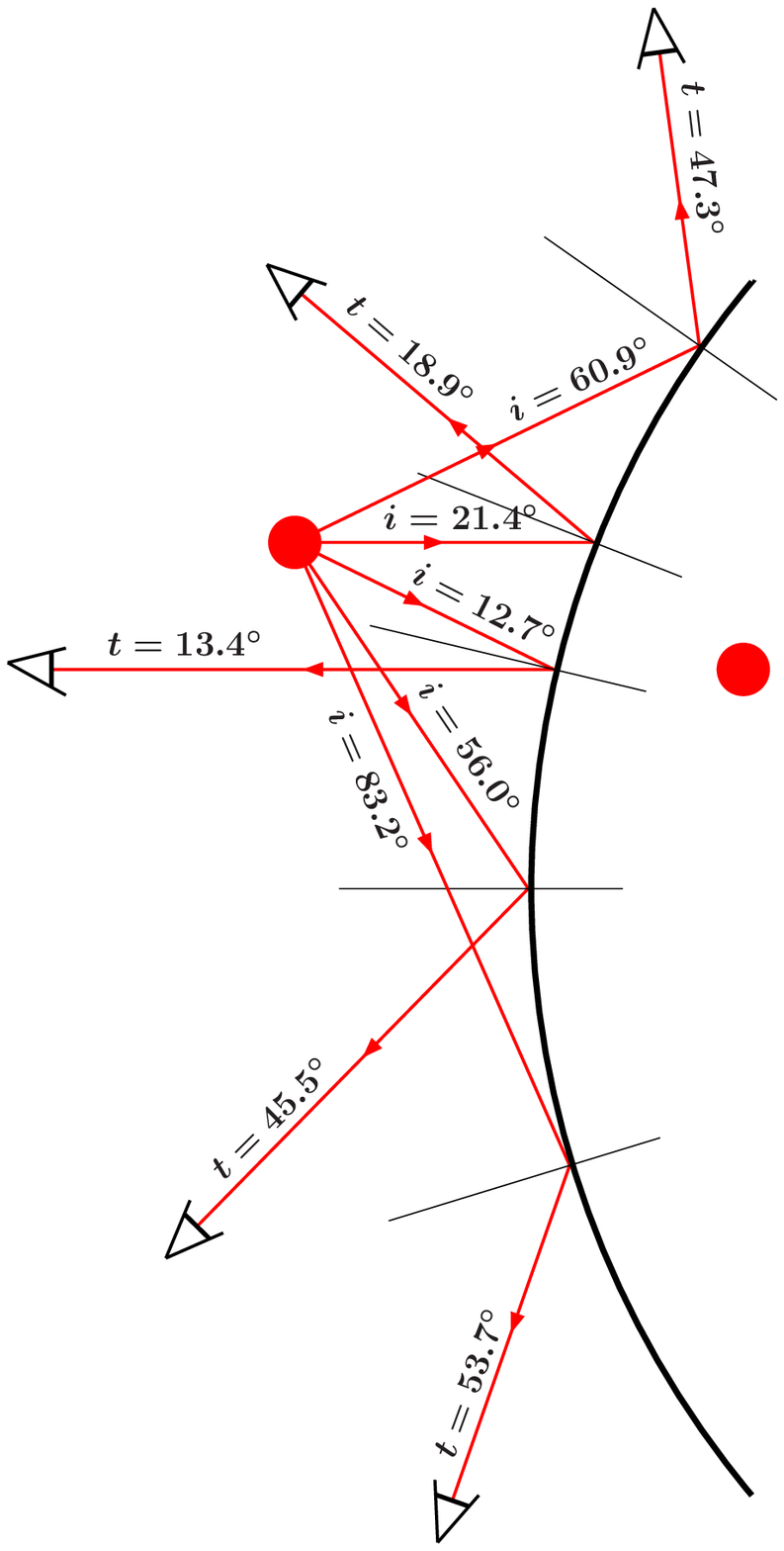}}} \\
\hspace{110pt}(a) & \hspace{80pt}(b) \\
\hline
\end{tabular}
\end{center}
\caption[]{\it 
The image of the light source is assumed at the intersection of the directions
in which observers 2 and 3 are supposed to receive light rays
originating from the light source and reflected by the mirror.
The light ray for observer 2 seems to come from the direction of the
supposed focus $F$ on
the supposed principal axis of the mirror, since
the incident light ray, from the light source to the mirror, is parallel
to the supposed principal axis.
Similarly, the light ray for observer 3 is reflected in the direction parallel
to the supposed principal axis, since it is emitted by the light source
in the direction of the supposed focus.
In the right-hand figure we show the incident and reflection angles
with respect to the local normal direction
at the various vertices.
}
\label{w1}
\end{figure}

However, ideal lenses should not be confused with ideal spherical mirrors.
The ideal spherical mirror is a mirror in the form of a perfect sphere.
No simple rules can be applied to construct its images.
Nevertheless, almost all physics courses for high-school and undergraduate
students contain figures similar
to the one represented in Fig.~\ref{w1}a,
in order to demonstrate image formation in convex
(or concave) spherical mirrors.
In Fig.~\ref{w1}b we show the results of measuring
the incident $i$ and reflection $t$ angles at each of the vertices,
applying simple Euclidean geometry.
Being rather obvious for observer 5, we actually find that
none of the vertices respects relation (\ref{teqi}).
We must therefore conclude that Fig.~\ref{w1}a cannot be correct.

The latter conclusion is no surprise, of course.
Since a perfect spherical mirror is spherically symmetric,
any choice of principal axis is as good (or bad) as any other.
Descartes' formula for the position of an image
does not imply the existence of a focus.
It just provides a simple method to determine the place of the image
for an observer positioned right behind the light source
with respect to the centre of the sphere.
For observers which are just a very small angle away from that direction,
it constitutes a reasonable approximation.
The method given by Descartes can thus perfectly well be applied
to optical instruments,
where long tubes guarantee that the angles involved are small.
However, for the images observed in Christmas balls,
it is not a correct method to be applied.

In the following paragraphs, we shall show a simple method of how
to determine the images seen by each of the observers in Fig.~\ref{w1}.
We are well aware that similar demonstrations must have been presented
long ago.
But unfortunately, we have not found any references to such work
(see also Ref~\cite{MVBerry}).
Since, moreover, it seems to have been completely forgotten, we judged it
useful to present our material to a wider public than just our students.
\clearpage

\section{Reflection.}

Let us consider a pointlike light source and a spherical mirror.
We assume that the eye lens is small enough to be considered pointlike
as well.  The relevant light rays must all be in the plane determined by
the centre of the sphere, the position of the light source, and the eye.
Consequently, we may study the subject in two dimensions.

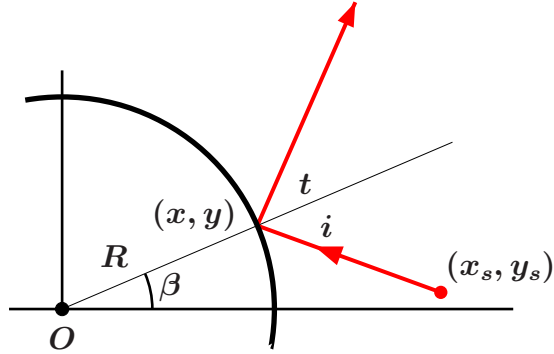
\begin{figure}[htbp]
\begin{center}
\input{figures/step1}
\end{center}
\caption[]{\it 
A light ray emitted by the light source at $(x_{s},y_{s})$
is reflected by the spherical mirror.
The incident angle $i$ and the reflection angle $t$
at the surface of the mirror are equal: $i=t\equiv\varphi$.
The vector connecting the coordinate origin to the vertex
at $(x,y)$ makes an angle $\beta$ with the horizontal axis.
}
\label{step1}
\end{figure}

We take the centre of the circle, which has radius $R$,
at the origin of the coordinate system.
In Fig.~\ref{step1} we show a light ray
emitted by a light source located at $(x_{s},y_{s})$,
and reflected by the spherically curved mirror at the vertex $(x,y)$.

Since we are essentially left with a two-dimensional problem,
we may perform the geometrical calculations by using complex numbers.
Thus, let us define the complex numbers
$r_{s}$ (object location) 
and $r$ (vertex location), i.e., 
\begin{equation}
{r_{s}}\; =\; x_{s}+iy_{s}
\;\;\;\xrm{and}\;\;\;
r\; =\; x+iy\; =\; R\; e^\x{i\beta}
\;\;\; .
\label{complex}
\end{equation}
The vector ${r_{s}}-r$ points from the vertex
to the light source (object) along the incident light ray.
Furthermore, the angle $i=t\equiv\varphi$ (see Fig.~\ref{step1})
is defined as the angle between $r_{s}-r$ and $r$.
Consequently,
\begin{equation}
\fnd{r}{{r_{s}}-r}\;
=\;\abs{\fnd{r}{{r_{s}}-r}}\; e^{i\varphi}
\;\;\;\;\xrm{or}\;\;\;\;
e^{i\varphi}\;
=\;\abs{\fnd{{r_{s}}-r}{r}}\fnd{r}{{r_{s}}-r}
\;\;\; .
\label{aphi}
\end{equation}
When by using Eq.~(\ref{aphi})
we rotate the vector ${r_{s}}-r$ over an angle $2\varphi$,
i.e.,
\begin{equation}
({r_{s}}-r)e^{2i\varphi}\; =\;
({r_{s}}-r)\left(\abs{\fnd{{r_{s}}-r}{r}}\;
\fnd{r}{{r_{s}}-r}\right)^{2}\; =\;
\fnd{{r_{s}}^{\ast}-r^{\ast}}{r^{\ast}}r
\;\;\; ,
\label{amreqbmr}
\end{equation}
then we obtain a vector in the direction of the reflected light ray.
Using now Eq.~(\ref{amreqbmr}), we deduce the following expression for
the straight line which coincides with the reflected light ray
($\lambda$ real):
\begin{equation}
s(\lambda )\;
=\; r\; +\;\lambda\fnd{{r_{s}}^{\ast}-r^{\ast}}{r^{\ast}}r
\;\;\; .
\label{reflected_ray}
\end{equation}
\clearpage

\section{The image}

\begin{figure}[htbp]
\centerline{\scalebox{0.5}{\includegraphics{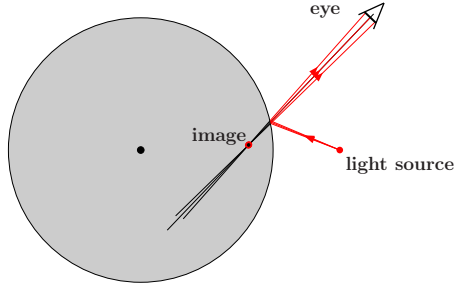}}}
\caption[]{\it 
Light rays which are reflected in nearby directions,
intersect at nearly the same location.
}
\label{image}
\end{figure}

The small bundle of light rays which strikes the eye lens
has a very small area of convergence,
as can be understood from Fig.~\ref{image}. 
This small area is where the eye supposes the light rays are stemming from,
hence the image of the light source.

By varying the angle $\beta$ in Fig.~\ref{step1}, we obtain different
trajectories for light rays originating from the same light source
at $r_{s}$.
We are here interested in the intersection of the straight lines
that coincide with the reflected rays of two different suchlike trajectories.
We indicate their vertices by $r$ and $r'{=r+\Delta r}$,
respectively.
We define the image of the light source, for observers in the direction
characterised by $r$, as the point of intersection
of the two trajectories,
in the limit of vanishing $\abs{\Delta r}$.
In the Appendix it is shown how one obtains thus,
the position $r_{i}$
of the image of the light source.
From Eq.~(\ref{Arimage}) one easily deduces the expression
\begin{equation}
r_{i}\; =\;
r\;\fnd{3\abs{{r_{s}}}^{2}-{r_{s}}^{\ast}
{r_{s}}^{\ast}\fnd{r}{r^{\ast}}-2{r_{s}}r^{\ast}}
{4\abs{{r_{s}}}^{2}-3{r_{s}}r^{\ast}
-3{r_{s}}^{\ast}r+2\abs{r}^{2}}
\;\;\; .
\label{rimage}
\end{equation}
The denominator in this fraction is real.
Consequently, it is an easy task to extract the $x$ and $y$
components of the image position.
One gets
\begin{eqnarray}
x_{i} & = &
R\;\fnd
{
x_{s}^{2}\cos (\beta )\left[ 1+2\sin^{2}(\beta )\right]
-x_{s}y_{s}\sin (3\beta )+2y_{s}^{2}\cos^{3}(\beta )-Rx_{s}
}
{R^{2}+2x_{s}^{2}+2y_{s}^{2}
-3R\left[ x_{s}\cos (\beta )+y_{s}\sin (\beta )\right]
}
\;\;\;\;,
\label{xyI}\\ [10pt]
y_{i} & = &
R\;\fnd{2x_{s}^{2}\sin^{3}(\beta )+
y_{s}^{2}\sin (\beta )\left[ 1+2\cos^{2}(\beta )\right]-Ry_{s}
+x_{s}y_{s}\cos (\beta )\left[ 1-4\sin^{2}(\beta )\right]}
{R^{2}+2x_{s}^{2}+2y_{s}^{2}
-3R\left[ x_{s}\cos (\beta )+y_{s}\sin (\beta )\right]}
\;\;\; .
\nonumber
\end{eqnarray}
\clearpage

\begin{figure}[htbp]
\begin{center}
\begin{tabular}{|c|c|}
\hline & \\
\input{figures/im1} & \input{figures/im2}
 \\ \hline
\input{figures/im3} & \input{figures/im4}
 \\ \hline
\end{tabular}
\end{center}
\caption[]{\it Some different situations.
The locations of the images are indicated by red dots
labelled $i$, at the ---
in general not exactly coinciding --- intersections of the three light rays.
The coordinates of these red dots follow from Eq.~(\ref{xyI}).}
\label{Imbol}
\end{figure}
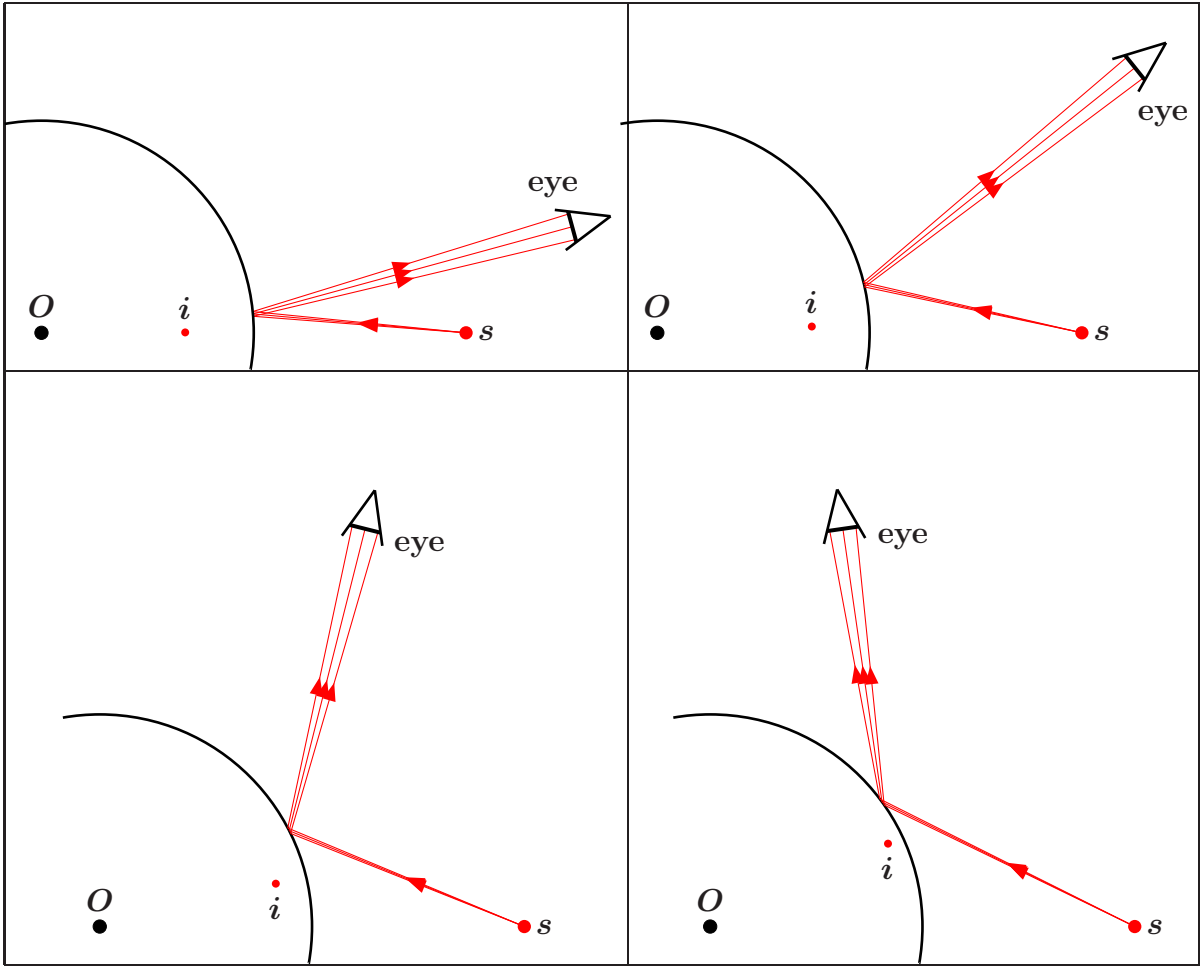

In Fig.~\ref{Imbol} we show some pictures.
We indicate the positions of the light source, the eye, and the
centre of the spherical mirror.
Three different light rays are reflected on the surface of the sphere
and then reach the eye.
We assume the location of the image to lie where these rays intersect
one another.
A small red dot labelled by $i$ indicates the position of the image
for the central direction, which follows from 
Eq.~(\ref{xyI}).
The image in the upper-left picture of Fig.~\ref{Imbol} can approximately
be reconstructed from Descartes' formula.
But, the other three images come obviously at different locations.

The reason that we see a ``sharp'' image of the pointlike light source
is due to the smallness of our eye lens.
More accurately, it stems from the ratio of the diameter of the
pupil of the eye and the curvature radius $R$ of the spherical mirror.
Hence, perfectly sharp images are only to be expected for an infinite
radius of curvature, which just represents a flat mirror.
Nevertheless, for large spherical mirrors, the resolution of the virtual image
observed by the eye is good enough to be considered sharp.
However, its {\it location} depends on the position of the eye.
\clearpage

For a light source on the $x$ axis ($y_{s}=0$) and for
$\beta =0$, we obtain for the image location
\begin{equation}
x_{i}\; =\; R\;\fnd{x_{s}}{2x_{s}-R}
\;\;\;\;\xrm{and}\;\;\;\;
y_{i}\; =\; 0
\;\;\; .
\label{bcentral}
\end{equation}
Moreover, one then finds
\begin{equation}
\fnd{1}{x_{s}}\; +\;\fnd{1}{x_{i}}\; =\;
\fnd{1}{x_{s}}\; +\;\fnd{2x_{s}-R}{Rx_{s}}\; =\;\fnd{2}{R}
\;\;\; ,
\label{Descartes}
\end{equation}
which is Descartes' formula for spherical mirrors.
\clearpage

\section{The angle \bm{\beta}}

The question is how to determine the angle $\beta$
in the case that the positions of the light source $(x_{s},y_{s})$
and the eye $(x_{0},y_{0})$ are given.

Let us define the complex number $r_{o}$
for the coordinates of the eye lens,
i.e.,
\begin{equation}
{r_{o}}\; =\; x_{o}+iy_{o}
\;\;\; .
\label{bcomplex}
\end{equation}
From Fig.~\ref{step1}, and in analogy with Eq.~(\ref{aphi}),
we obtain 
\begin{equation}
{r_{o}}-r\;
=\;\abs{{r_{o}}-r}\; e^{i(\beta +\varphi )}
\;\;\;\xrm{and}\;\;\;
\fnd{{r_{o}}-r}{r}\;
=\;\abs{\fnd{{r_{o}}-r}{r}}\; e^{i\varphi}
\;\;\; .
\label{bphi}
\end{equation}
Combining Eqs.~(\ref{aphi}) and (\ref{bphi}) gives
\begin{equation}
\abs{\fnd{{r_{s}}-r}{r}}\;\fnd{r}{{r_{s}}-r}\; =\;
e^{i\varphi}\; =\;
\abs{\fnd{r}{{r_{o}}-r}}\;\fnd{{r_{o}}-r}{r}
\;\;\; ,
\label{eialfa}
\end{equation}
or
\begin{equation}
\fnd{({r_{s}}-r)({r_{o}}-r)}{r^{2}}\; =\;
\abs{\fnd{({r_{s}}-r)({r_{o}}-r)}{r^{2}}}
\;\;\; .
\label{central}
\end{equation}
The right-hand side of Eq.~(\ref{central}) is real,
whence for the imaginary part of the left-hand side we conclude
\begin{equation}
\fnd{({r_{s}}-r)({r_{o}}-r)}{r^{2}}\; -\;
\left(\fnd{({r_{s}}-r)({r_{o}}-r)}{r^{2}}\right)^{\ast}
\; =\; 0
\;\;\; ,
\label{Imcentral}
\end{equation}
which is equivalent to the fourth-order equation
\begin{equation}
\fnd{{r_{s}}^{\ast}{r_{o}}^{\ast}}{R^{2}}
\;\left( e^{i\beta}\right)^{4}
-\fnd{{r_{s}}^{\ast}+{r_{o}}^{\ast}}{R}
\;\left( e^{i\beta}\right)^{3}
+\fnd{{r_{s}}+{r_{o}}}{R}\;\left( e^{i\beta}\right)
-\fnd{{r_{s}}{r_{o}}}{R^{2}}\; =\; 0
\;\;\; .
\label{Im4central}
\end{equation}
\clearpage

\subsection{The solutions to Eq.~(\ref{Im4central})}

The biquadratic Eq.~(\ref{Im4central}) is in the literature known
as the {\it Billiard Problem of al-Hasan}, named after the Arab scientist
{\it Abu Ali al-Hasan ibn al-Haytham} \/(965--1040 A.D.)
\cite{Wolfram,Fisher47}.
It corresponds to the following problem, which had already been formulated
in 150 A.D.\ by the Greek scientist {\it Ptolemy}:
``Find, for arbitrary initial positions of the red ball and the white ball, 
the point on the edge of a circular billiard table
at which the white ball must be aimed in order to carom once off the edge
and collide head-on with the red ball''.

\mbox{From} the definitions of $r_{s}$,
$r_{o}$, $r$,
and the subsequent definition of $\beta$ in Eq.~(\ref{complex}),
it may be clear that reflection in the interior of the circle representing
the cross section of the spherical mirror
can be studied in exactly the same way as reflection in the exterior,
by considering $r-{r_{s}}$ instead of ${r_{s}}-r$,
and $r-{r_{o}}$ instead of ${r_{o}}-r$,
which leads to exactly the same Eq.~(\ref{Im4central}).

In general, there are four solutions $z_{1}$, $z_{2}$, $z_{3}$, and $z_{4}$
to the equation
\begin{equation}
\fnd{{r_{s}}^{\ast}{r_{o}}^{\ast}}{R^{2}}\; z^{4}
\;-\;\fnd{{r_{s}}^{\ast}+{r_{o}}^{\ast}}{R}\; z^{3}
\;+\;\fnd{{r_{s}}+{r_{o}}}{R}\; z
\;-\;\fnd{{r_{s}}{r_{o}}}{R^{2}}\; =\; 0
\;\;\; .
\label{4poly}
\end{equation}
However, when $\abs{z}\neq 1$, the solution to Eq.~(\ref{4poly}) is not a
physical solution to Eq.~(\ref{Im4central}).
Although we have not studied the situation exhaustively,
we find that generally there are two solutions to Eq.~(\ref{Im4central}),
but sometimes there are four (see also Ref.~\cite{Drexler}),
depending on the complex parameters $r_{s}$
and $r_{o}$.
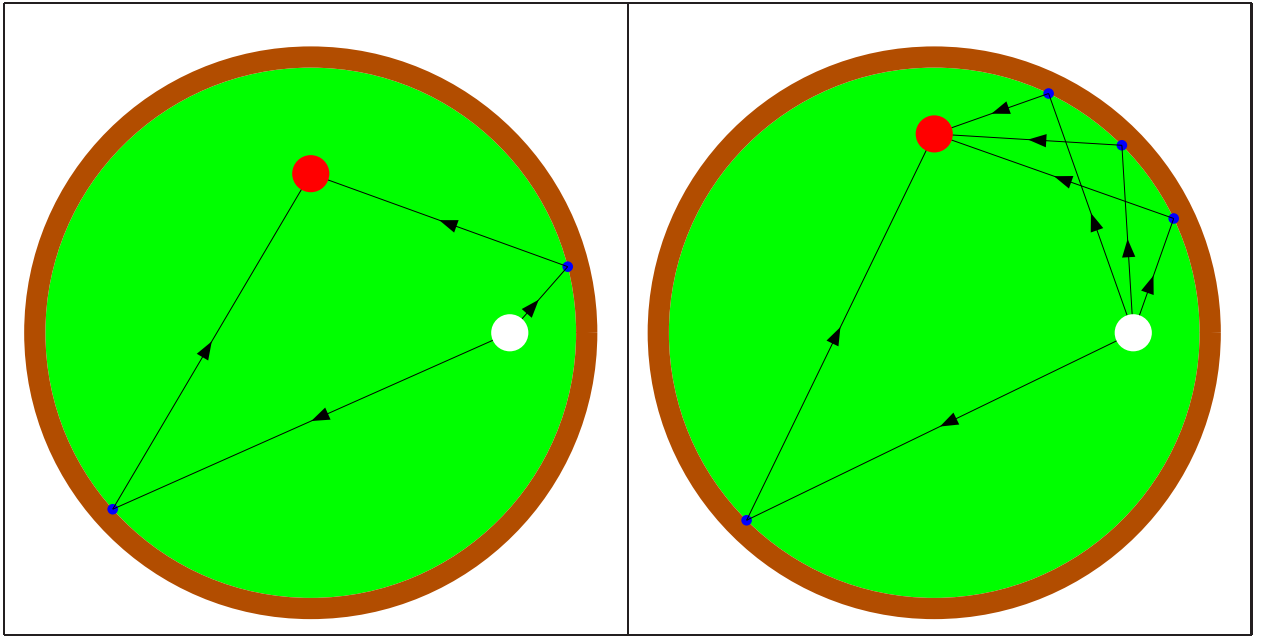
\begin{figure}[htbp]
\begin{center}
\begin{tabular}{|c|c|}
\hline & \\
\input{figures/bil1}
 &
\input{figures/bil2}
\\ \hline
\end{tabular}
\end{center}
\caption[]{\it Two different situations for the initial positions of the two
billiard balls.
In the left-hand picture, we have an example where there are only two
solutions to Eq.~(\ref{Im4central}).
The right-hand picture shows an example where, by just selecting a
different position of the red ball, we find four
solutions.
}
\label{Billiard}
\end{figure}
The different cases, corresponding to ${r_{s}}=0$,
${r_{o}}=0$, ${r_{s}}={r_{o}}=0$,
$\abs{{r_{s}}}=R$,
$\abs{{r_{o}}}=R$,
and $\abs{{r_{s}}}=\abs{{r_{o}}}=R$,
must be studied separately.
In Fig.~\ref{Billiard} we show two examples for a circular billiard.
\clearpage

\section{Deformation}

\begin{figure}[htbp]
\begin{center}
\begin{tabular}{|p{180pt}|}
\hline \\
\centerline{\scalebox{0.5}{\includegraphics{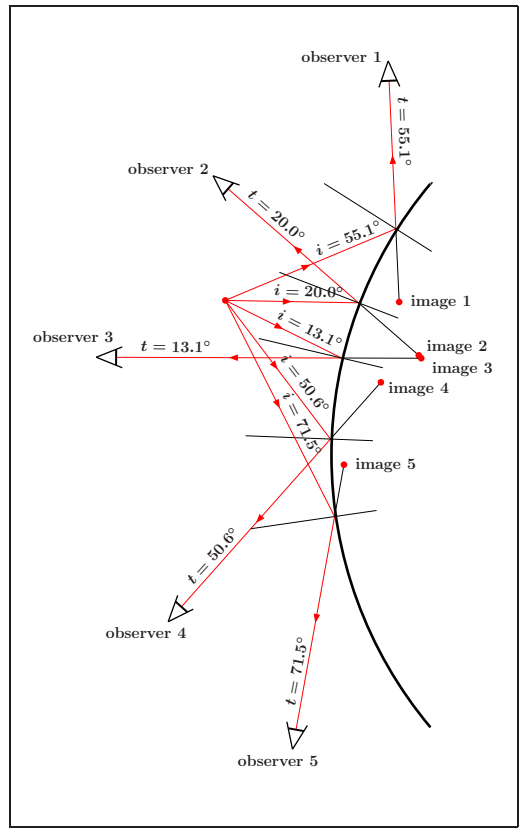}}} \\
\hline
\end{tabular}
\end{center}
\caption[]{\it The locations of the various images as seen by each of the five
observers introduced in Fig.~\ref{w1}.
We also indicate the angles of incidence and reflection, in order
to make sure that they are equal.}
\label{c1}
\end{figure}

Now that we have solved the problem of finding the vertex on the curved
mirror where the light ray passing from the object to the eye is reflected,
we may construct the images for the situation shown in Fig.~\ref{w1}.
The result is depicted in Fig.~\ref{c1}.
\clearpage

\begin{figure}[htbp]
\begin{center}
\begin{tabular}{|p{90pt}|p{100pt}|p{90pt}|p{80pt}|}
\hline & & & \\
\centerline{\scalebox{0.5}{\includegraphics{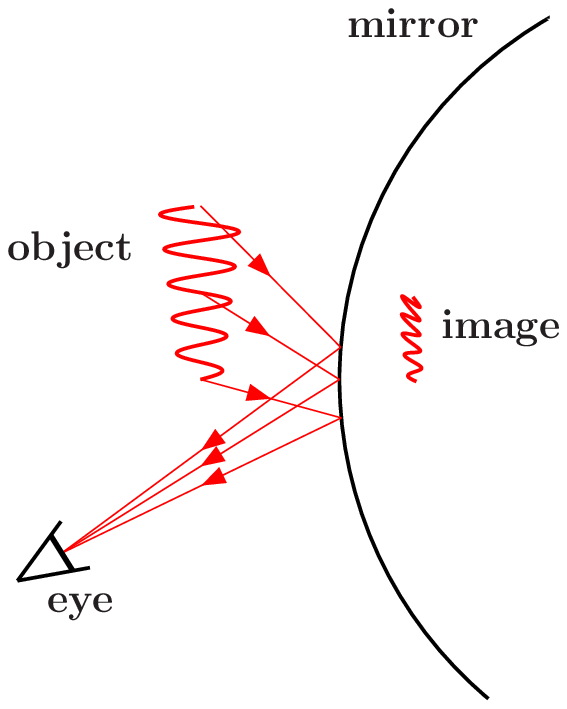}}} &
\centerline{\scalebox{0.5}{\includegraphics{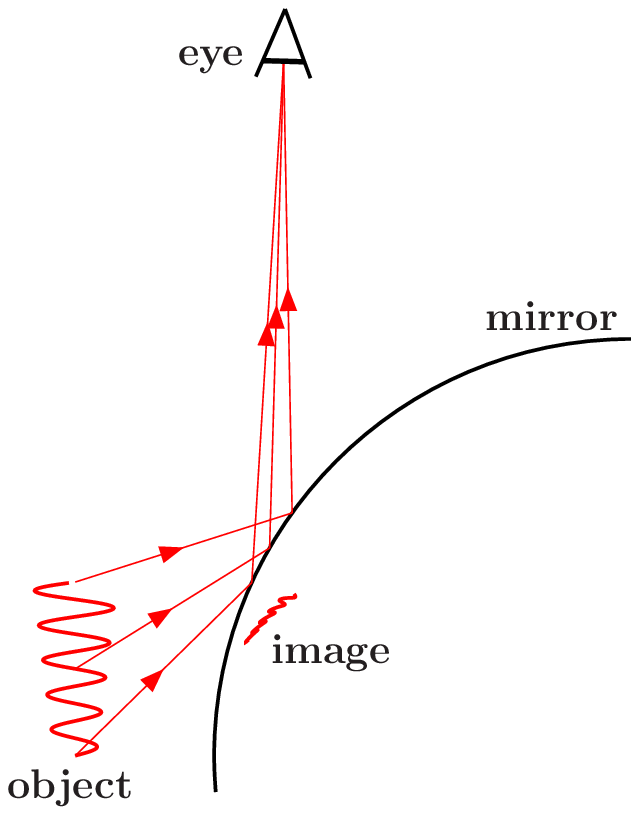}}} &
\centerline{\scalebox{0.5}{\includegraphics{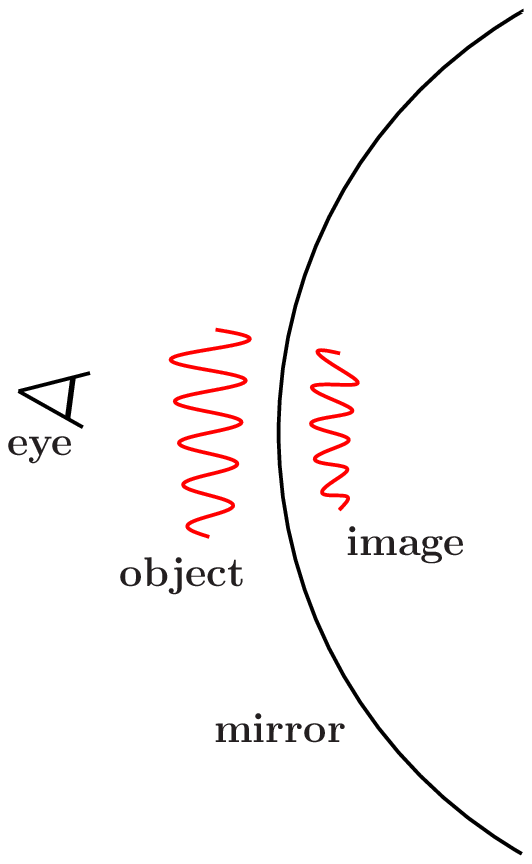}}} &
\centerline{\scalebox{0.5}{\includegraphics{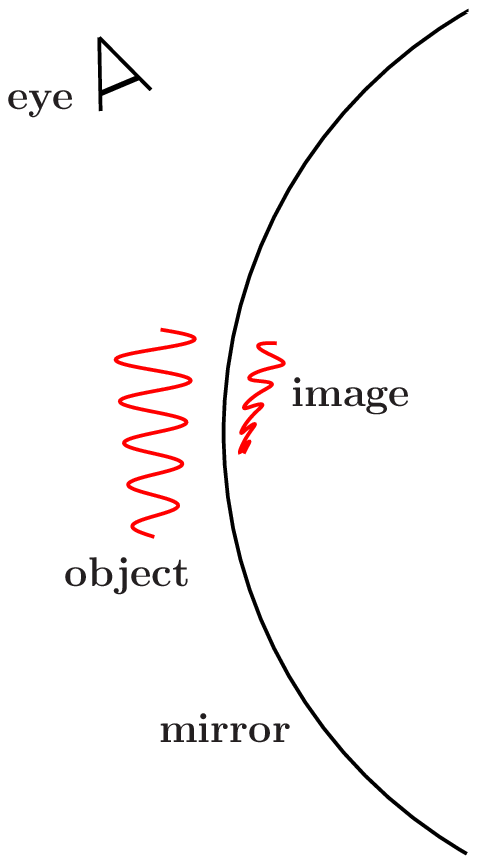}}}
\\ \hline
\end{tabular}
\end{center}
\caption[]{\it 
Images of an extended object, as seen in a Christmas ball
from different angles.
}
\label{Defbol}
\end{figure}

Furthermore, we may now construct the images of extended objects,
and study their deformation.
In Fig.~\ref{Defbol}, we give a few simple examples.
\clearpage

\section{Conclusions}

The complete resolution of the Billiard Problem (\ref{Im4central}),
posed by al-Hasan in the context of what nowadays is called geometric
optics, had to wait for many centuries.
First it was necessary to derive an analytic form for the
solutions to biquadratic equations like Eq.~(\ref{4poly}).
Apparently found by Ferrari, it was published for the first time
in Cardano's ``Ars Magna'' in 1545.
The first one to subsequently solve the Billiard Problem of al-Hasan
was Christiaan Huyghens.

The problem would probably have ended up in the history books,
since ``once solved, forever solved''.
However, for optical equipment, where the angles of reflection
$\varphi$ are small, Descartes formulated an approximation,
amounting to basically linear equations.
Anything that has larger reflection angles is considered an
{\it aberration} \/in his philosophy.
Ever since, a whole generation of physicists emerged completely unaware
of al-Hasan's Billiard Problem, who systematically applied Descartes'
approximation also to large angles $\varphi$, thus arriving at totally
wrong conclusions. Therefore, we should rub up al-Hasan's Billiard Problem,
and reintroduce it in our optics courses, so that
new generations can fully enjoy the perfectly sharp images in Christmas balls,
without feeling uneasy for having been told that such wonderful images
should be considered aberrations.

For undergraduate students the full subject can be treated,
involving some simple computer programming.
One can either choose a point on the spherical mirror, and construct
the image by the use of formulae (\ref{xyI}) and from the direction
of the reflected light ray, or one can select the position of the eye pupil,
and determine the vertex by means of Eq.~(\ref{Im4central}).
However, for youngsters at high schools, we recommend to only deal with it
qualitatively, using examples like Figs.~\ref{image} and \ref{c1}.

\section*{Acknowledgments}

We are grateful for many useful discussions with our colleagues,
in particular with Alex Blin, Brigitte Hiller,
Jo\~{a}o da Provid\^{e}ncia Santar\'{e}m e Costa,
Francisco Gil and Constan\c{c}a da Pro\-vi\-d\^{e}n\-cia,
as well as with Jo\~{a}o Paulo Fonseca from the
Escola Secund\'{a}ria de Tondela.

\clearpage

\appendix

\section{The point of convergence for adjacent reflected light rays}

We consider here two reflected light rays,
characterised by two different vertices $r$ and $r'$. 
In accordance with formula (\ref{reflected_ray}), we may represent
the two straight lines coinciding with the two reflected rays
by ($\lambda$, $\lambda '$ real)
\begin{equation}
s(\lambda )\; =\; r\; +\;\lambda\fnd{r_{s}^{\ast}-r^{\ast}}{r^{\ast}}r
\;\;\;\;\xrm{and}\;\;\;\;
s'(\lambda ')\;
=\; r'\; +\;\lambda '\fnd{r_{s}^{\ast}-{r'}^{\ast}}{{r'}^{\ast}}r'
\;\;\; .
\label{ray12}
\end{equation}
The intersection of the two trajectories follows from 
\begin{equation}
s(\lambda )\; =\; s'(\lambda ')
\;\;\; ,
\label{intersect1}
\end{equation}
which, since $\lambda$ and $\lambda '$ are real, actually corresponds to two
equations, one for the real part and one for the imaginary part,
namely
\begin{equation}
s(\lambda )\;\pm\; s^{\ast}(\lambda )\; =\;
s'(\lambda ')\;\pm\; {s'}^{\ast}(\lambda ')
\;\;\; .
\label{intersect2}
\end{equation}
The system of equations (\ref{intersect2}) can be solved
for $\lambda$ and $\lambda '$.

In order to avoid lengthy formulas, we define
\begin{equation}
A\, =\,\fnd{r_{s}^{\ast}-r^{\ast}}{r^{\ast}}r
\;\;\;\;\xrm{and}\;\;\;\;
{A'}\, =\,\fnd{r_{s}^{\ast}-{r'}^{\ast}}{{r'}^{\ast}}r'
\;\;\; .
\label{ABdef}
\end{equation}
Using definitions (\ref{ABdef}) and formula (\ref{ray12}), we obtain
\begin{displaymath}
s(\lambda )\pm s^{\ast}(\lambda ) =
r\pm r^{\ast} +\lambda\left( A\pm A^{\ast}\right)
\;\;\;\;\xrm{and}\;\;\;\;
\end{displaymath}
\begin{equation}
s'(\lambda' )\pm{s'}^{\ast}(\lambda' ) =
r'\pm{r'}^{\ast}+\lambda'\left( {A'}\pm {A'}^{\ast}\right)
\;\;\; .
\label{spmsstar}
\end{equation}
Inserting the result (\ref{spmsstar}) into the two relations
(\ref{intersect2}) gives the two equations
\begin{equation}
\left( A\,\pm\, A^{\ast}\right)\,\lambda\, -\,
\left( {A'}\,\pm\, {A'}^{\ast}\right)\,\lambda'\, =\,
-\, r\,\mp\, r^{\ast}\, +\,{r'}\,\pm\,{r'}^{\ast} ,
\label{intersect3}
\end{equation}
which can be cast in matrix form
and are readily solved by
\begin{eqnarray}
\left(\begin{array}{c} \lambda\\ [10pt] {\lambda'}\end{array}\right)
 & = &
\left(\begin{array}{cc}
A+A^{\ast}\;\; & -{A'}-{A'}^{\ast}\\ [10pt]
A-A^{\ast}\;\; & -{A'}+{A'}^{\ast}
\end{array}\right)^{-1}\,
\left(\begin{array}{c}
-r-r^{\ast}+{r'}+{r'}^{\ast}\\ [10pt] 
-r+r^{\ast}+{r'}-{r'}^{\ast}
\end{array}\right)
\nonumber\\ [10pt] & = &
\fndrs{19pt}{
\left(\begin{array}{c}
{A'}\left( r^{\ast}-{r'}^{\ast}\right)\, -\,
{A'}^{\ast}(r-r')
\\ [10pt] 
A\left( r^{\ast}-{r'}^{\ast}\right)\, -\,
A^{\ast}(r-r')
\end{array}\right)
}
{-5pt}
{A{A'}^{\ast}\, -\, A^{\ast}{A'}}
\;\;\; .
\label{lambdalambdaprime}
\end{eqnarray}
Next, we take $r'\; =\; r\; +\;\Delta r$
and expand the various terms of Eq.~(\ref{lambdalambdaprime})
to first order in $\Delta r$ and
$\Delta r^{\ast}$.
In the limit $\abs{\Delta r}\downarrow 0$, one has moreover
\begin{equation}
\fnd{\Delta r^{\ast}}{\Delta r}\;\longrightarrow\;
-\;\fnd{r^{\ast}}{r}
\;\;\; .
\label{drsodr}
\end{equation}
This can most easily be understood, when one, using Eq.~(\ref{complex}),
defines
\begin{displaymath}
r+\Delta r=Re^\x{i(\beta +\Delta\beta)}\approx
Re^\x{i\beta}(1+i\Delta\beta)
\;\;\; ,
\end{displaymath}
hence, to lowest order in $\Delta\beta$, one finds 
$\Delta r=iRe^\x{i\beta}\Delta\beta$.
Consequently, since $R$, $\beta$ and $\Delta\beta$ are real,
\begin{displaymath}
\fnd{\left\{\Delta r\right\}^{\ast}}{\Delta r}\;\longrightarrow\;
-\fnd{Re^\x{-i\beta}}{Re^\x{i\beta}}=-\fnd{r^{\ast}}{r}
\;\;\; .
\end{displaymath}
Using, moreover, the definitions (\ref{ABdef}), we determine,
in the limit $\abs{\Delta r}\downarrow 0$,
\begin{eqnarray*}
{A'} & = & \fnd{a^{\ast}-r^{\ast}-\Delta r^{\ast}}{r^{\ast}+\Delta r^{\ast}}
(r+\Delta r)
\approx
A\left( 1+\fnd{\Delta r}{r}-\fnd{\Delta r^{\ast}}{r^{\ast}}\right)
-\fnd{r}{r^{\ast}}\Delta r^{\ast}
\\ [10pt] & \approx &
A+\fnd{\Delta r}{r}\left\{
A\left( 1-\fnd{r}{r^{\ast}}\fnd{\Delta r^{\ast}}{\Delta r}\right)
-r\fnd{r}{r^{\ast}}\fnd{\Delta r^{\ast}}{\Delta r}\right\}
\longrightarrow
A+\fnd{\Delta r}{r}( 2A+r)
\; ,
\end{eqnarray*}
and similarly,
\begin{equation}
A{A'}^{\ast}-A^{\ast}{A'}\longrightarrow\fnd{\Delta r}{r}
\left( -4\abs{A}^{2}-Ar^{\ast}-A^{\ast}r\right)
\;\;\; ,
\label{Detover2}
\end{equation}
\begin{equation}
{A'}\left( r^{\ast}-{r'}^{\ast}\right) -{A'}^{\ast}(r-r')\longrightarrow
\fnd{\Delta r}{r}\left( Ar^{\ast}+A^{\ast}r\right)
\;\;\; .
\label{Numerator}
\end{equation}

On substitution of the results (\ref{Detover2}) and (\ref{Numerator})
into expressions (\ref{lambdalambdaprime}),
we obtain for
the parameter $\lambda$ at the point of
intersection,
in the limit of vanishing $\abs{\Delta r}$, the result
\begin{equation}
\lambda =-\fnd{Ar^{\ast}+A^{\ast}r}
{4\abs{A}^{2}+Ar^{\ast}+A^{\ast}r}
\;\;\; .
\label{Lambda}
\end{equation}

Finally, using Eq.~(\ref{Lambda}),
we determine the point of intersection, $r_{i}$,
of the two straight lines (\ref{ray12})
in the limit of vanishing $\abs{\Delta r}$.
\begin{equation}
r_{i}\; =\; r+\lambda A =
\fnd{4\abs{A}^{2}r+(Ar^{\ast}+A^{\ast}r)(r-A)}
{4\abs{A}^{2}+Ar^{\ast}+A^{\ast}r}
\;\;\; .
\label{Arimage}
\end{equation}

The variable $\lambda$ indicates how deep the ``image'' is
below the surface of the spherical mirror,
measured from the mirror surface, along the direction of reflection,
to the point where the small bundle of reflected rays seems to emerge.
Alternatively, for practical purposes,
one could determine this ``depth'' from
\begin{displaymath}
\fndrs{4pt}{1}{-5pt}{\abs{\lambda A}}\; =\;
\fndrs{4pt}{2}{-5pt}{R\cos (\varphi )}\; +\;
\fndrs{4pt}{1}{-5pt}{\abs{r_{s}-r}}
\;\;\; . 
\end{displaymath}
One measures with a ruler, the lengths of $\abs{\vec{r}_{s}-\vec{r}\,}$
and $R\cos (\varphi )$, the latter equals
half the distance from the vertex to the other intersection of the
mirror and the line which coincides with the reflected light ray,
and performs the above calculation.
For small angles $\varphi$, one obtains the image location of Descartes,
whereas, for larger angles the location of the image is different.
For $\varphi = 90^{\circ}$, one obtains zero depth.
\end{document}

%% file: figures/step1.tex
\begin{picture}(194,125)(-20,-10)
\special{"
1.0 0.0 0.0 setrgbcolor
0.0 setlinewidth
newpath 145.21 6.29 moveto 142.71 6.29 2.5 0 360 arc closepath
fill stroke}%
\special{"
1.0 0.0 0.0 setrgbcolor
1.5 setlinewidth
newpath  142.71 6.29 moveto   73.55  31.46 lineto stroke}%
\special{"
1.0 0.0 0.0 setrgbcolor
0.0 setlinewidth newpath  95.16 23.59 moveto 103.56 16.55 lineto
106.13  23.60 lineto closepath
fill stroke}%
\special{"
0.2 setlinewidth
newpath     0.00    0.00 moveto  147.11   62.93 lineto stroke}%
\special{"
1.0 0.0 0.0 setrgbcolor
1.5 setlinewidth
newpath   73.55   31.46 moveto  110.50  115.71 lineto stroke}%
\special{"
1.0 0.0 0.0 setrgbcolor
0.0 setlinewidth newpath 110.50  115.71 moveto 103.85  108.75 lineto
109.88  106.11 lineto closepath
fill stroke}%
\special{"
1.0 setlinewidth
newpath -20 0 moveto 170 0 lineto stroke}%
\special{"
1.0 setlinewidth
newpath 0 0 moveto 0 90 lineto stroke}%
\special{"
2.0 setlinewidth
newpath 78.78  -13.89 moveto 0 0 80 -10 100 arc stroke}%
\special{"
0.0 setlinewidth
newpath 2.5 0 moveto 0 0 2.5 0 360 arc closepath
gsave 0.0 setgray fill grestore stroke}%
\special{"
1.0 setlinewidth
newpath 34 0 moveto 0 0 34 0 23 arc stroke}%
\put(0,-7){\makebox(0,0)[tc]{\bm{O}}}
\put(145,10){\makebox(0,0)[bl]{\bm{(x_{s},y_{s})}}}
\put(66,35.5){\makebox(0,0)[rc]{\bm{(x,y)}}}
\put(25,16){\makebox(0,0)[rb]{\bm{R}}}
\put(97.5,31.5){\makebox(0,0)[lc]{\bm{i}}}
\put(89.5,47.5){\makebox(0,0)[lc]{\bm{t}}}
\put(45,3){\makebox(0,0)[rb]{\bm{\beta}}}
\end{picture}

%% file: figures/im1.tex
\begin{picture}(220,90)(-8,-10)
\special{"
0.0 setlinewidth
newpath 2.5 0 moveto 0 0 2.5 0 360 arc closepath
gsave 0.0 setgray fill grestore stroke}%
\special{"
    1.00    0.00    0.00 setrgbcolor
0.0 setlinewidth
newpath  162.50    0.00 moveto  160.00    0.00 2.5 0 360 arc closepath
fill stroke}%
\special{"
    1.00 setlinewidth
newpath  214.47   43.95 moveto  193.45   46.36 lineto stroke}%
\special{"
    1.00 setlinewidth
newpath  214.47   43.95 moveto  197.59   31.20 lineto stroke}%
\special{"
    1.00    0.00    0.00 setrgbcolor
    0.40 setlinewidth
newpath  200.00   40.00 moveto   79.68    7.12 lineto stroke}%
\special{"
    1.00    0.00    0.00 setrgbcolor
0.0 setlinewidth newpath 139.84   23.56 moveto 132.56   24.16 lineto
133.87   19.34 lineto closepath
fill stroke}%
\special{"
    1.00    0.00    0.00 setrgbcolor
    0.40 setlinewidth
newpath   79.68    7.12 moveto  160.00    0.00 lineto stroke}%
\special{"
    1.00    0.00    0.00 setrgbcolor
0.0 setlinewidth newpath 119.84    3.56 moveto 126.46    0.46 lineto
126.90    5.44 lineto closepath
fill stroke}%
\special{"
    1.00    0.00    0.00 setrgbcolor
    0.40 setlinewidth
newpath  198.68   44.82 moveto   79.60    8.01 lineto stroke}%
\special{"
    1.00    0.00    0.00 setrgbcolor
0.0 setlinewidth newpath 139.14   26.42 moveto 131.84   26.78 lineto
133.32   22.00 lineto closepath
fill stroke}%
\special{"
    1.00    0.00    0.00 setrgbcolor
    0.40 setlinewidth
newpath   79.60    8.01 moveto  160.00    0.00 lineto stroke}%
\special{"
    1.00    0.00    0.00 setrgbcolor
0.0 setlinewidth newpath 119.80    4.00 moveto 126.39    0.84 lineto
126.88    5.81 lineto closepath
fill stroke}%
\special{"
    1.00    0.00    0.00 setrgbcolor
    0.40 setlinewidth
newpath  201.32   35.18 moveto   79.76    6.24 lineto stroke}%
\special{"
    1.00    0.00    0.00 setrgbcolor
0.0 setlinewidth newpath 140.54   20.71 moveto 133.28   21.55 lineto
134.43   16.68 lineto closepath
fill stroke}%
\special{"
    1.00    0.00    0.00 setrgbcolor
    0.40 setlinewidth
newpath   79.76    6.24 moveto  160.00    0.00 lineto stroke}%
\special{"
    1.00    0.00    0.00 setrgbcolor
0.0 setlinewidth newpath 119.88    3.12 moveto 126.53    0.09 lineto
126.92    5.08 lineto closepath
fill stroke}%
\special{"
    1.00    0.00    0.00 setrgbcolor
0.0 setlinewidth
newpath   55.66    0.15 moveto   54.16    0.15 1.5 0 360 arc closepath
fill stroke}%
\special{"
    1.50 setlinewidth
newpath  198.42   45.79 moveto  201.58   34.21 lineto stroke}%
\special{"
1.0 setlinewidth
newpath 78.78  -13.89 moveto 0 0 80 -10 100 arc stroke}%
\put(0,6){\makebox(0,0)[bc]{\bm{O}}}
\put(165,0){\makebox(0,0)[lc]{\bm{s}}}
\put(54.16,5.15){\makebox(0,0)[bc]{\bm{i}}}
\put(193.45,51.36){\makebox(0,0)[bc]{\bf eye}}
\end{picture}

%% file: figures/im2.tex
\begin{picture}(200,120)(-5,-10)
\special{"
0.0 setlinewidth
newpath 2.5 0 moveto 0 0 2.5 0 360 arc closepath
gsave 0.0 setgray fill grestore stroke}%
\special{"
1.0 setlinewidth
newpath 78.78  -13.89 moveto 0 0 80 -10 100 arc stroke}%
\special{"
    1.00    0.00    0.00 setrgbcolor
0.0 setlinewidth
newpath  162.50    0.00 moveto  160.00    0.00 2.5 0 360 arc closepath
fill stroke}%
\special{"
    1.00 setlinewidth
newpath  191.70  109.38 moveto  171.46  103.23 lineto stroke}%
\special{"
    1.00 setlinewidth
newpath  191.70  109.38 moveto  181.29   90.97 lineto stroke}%
\special{"
    1.00    0.00    0.00 setrgbcolor
    0.40 setlinewidth
newpath  180.00  100.00 moveto   77.91   18.16 lineto stroke}%
\special{"
    1.00    0.00    0.00 setrgbcolor
0.0 setlinewidth newpath 128.96   59.08 moveto 122.03   56.73 lineto
125.16   52.83 lineto closepath
fill stroke}%
\special{"
    1.00    0.00    0.00 setrgbcolor
    0.40 setlinewidth
newpath   77.91   18.16 moveto  160.00    0.00 lineto stroke}%
\special{"
    1.00    0.00    0.00 setrgbcolor
0.0 setlinewidth newpath 118.96    9.08 moveto 125.12    5.16 lineto
126.20   10.04 lineto closepath
fill stroke}%
\special{"
    1.00    0.00    0.00 setrgbcolor
    0.40 setlinewidth
newpath  176.87  103.90 moveto   77.70   19.04 lineto stroke}%
\special{"
    1.00    0.00    0.00 setrgbcolor
0.0 setlinewidth newpath 127.29   61.47 moveto 120.44   58.90 lineto
123.69   55.10 lineto closepath
fill stroke}%
\special{"
    1.00    0.00    0.00 setrgbcolor
    0.40 setlinewidth
newpath   77.70   19.04 moveto  160.00    0.00 lineto stroke}%
\special{"
    1.00    0.00    0.00 setrgbcolor
0.0 setlinewidth newpath 118.85    9.52 moveto 124.98    5.53 lineto
126.11   10.41 lineto closepath
fill stroke}%
\special{"
    1.00    0.00    0.00 setrgbcolor
    0.40 setlinewidth
newpath  183.13   96.10 moveto   78.11   17.29 lineto stroke}%
\special{"
    1.00    0.00    0.00 setrgbcolor
0.0 setlinewidth newpath 130.62   56.69 moveto 123.62   54.57 lineto
126.63   50.57 lineto closepath
fill stroke}%
\special{"
    1.00    0.00    0.00 setrgbcolor
    0.40 setlinewidth
newpath   78.11   17.29 moveto  160.00    0.00 lineto stroke}%
\special{"
    1.00    0.00    0.00 setrgbcolor
0.0 setlinewidth newpath 119.05    8.65 moveto 125.26    4.78 lineto
126.29    9.67 lineto closepath
fill stroke}%
\special{"
    1.00    0.00    0.00 setrgbcolor
0.0 setlinewidth
newpath   59.72    2.37 moveto   58.22    2.37 1.5 0 360 arc closepath
fill stroke}%
\special{"
    1.50 setlinewidth
newpath  176.25  104.68 moveto  183.75   95.32 lineto stroke}%
\put(0,6){\makebox(0,0)[bc]{\bm{O}}}
\put(165,0){\makebox(0,0)[lc]{\bm{s}}}
\put(58.22,7.37){\makebox(0,0)[bc]{\bm{i}}}
\put(191.29,85.97){\makebox(0,0)[tc]{\bf eye}}
\end{picture}

%% file: figures/im3.tex
\begin{picture}(160,180)(0,-10)
\special{"
0.0 setlinewidth
newpath 2.5 0 moveto 0 0 2.5 0 360 arc closepath
gsave 0.0 setgray fill grestore stroke}%
\special{"
1.0 setlinewidth
newpath 78.78  -13.89 moveto 0 0 80 -10 100 arc stroke}%
\special{"
    1.00    0.00    0.00 setrgbcolor
0.0 setlinewidth
newpath  162.50    0.00 moveto  160.00    0.00 2.5 0 360 arc closepath
fill stroke}%
\special{"
    1.00 setlinewidth
newpath  103.66  164.55 moveto   91.25  147.41 lineto stroke}%
\special{"
    1.00 setlinewidth
newpath  103.66  164.55 moveto  106.49  143.58 lineto stroke}%
\special{"
    1.00    0.00    0.00 setrgbcolor
    0.40 setlinewidth
newpath  100.00  150.00 moveto   71.37   36.14 lineto stroke}%
\special{"
    1.00    0.00    0.00 setrgbcolor
0.0 setlinewidth newpath  85.69   93.07 moveto  81.59   87.02 lineto
 86.44   85.80 lineto closepath
fill stroke}%
\special{"
    1.00    0.00    0.00 setrgbcolor
    0.40 setlinewidth
newpath   71.37   36.14 moveto  160.00    0.00 lineto stroke}%
\special{"
    1.00    0.00    0.00 setrgbcolor
0.0 setlinewidth newpath 115.69   18.07 moveto 121.10   13.16 lineto
122.99   17.79 lineto closepath
fill stroke}%
\special{"
    1.00    0.00    0.00 setrgbcolor
    0.40 setlinewidth
newpath   95.15  151.22 moveto   70.85   37.16 lineto stroke}%
\special{"
    1.00    0.00    0.00 setrgbcolor
0.0 setlinewidth newpath  83.00   94.19 moveto  79.12   87.99 lineto
 84.01   86.95 lineto closepath
fill stroke}%
\special{"
    1.00    0.00    0.00 setrgbcolor
    0.40 setlinewidth
newpath   70.85   37.16 moveto  160.00    0.00 lineto stroke}%
\special{"
    1.00    0.00    0.00 setrgbcolor
0.0 setlinewidth newpath 115.42   18.58 moveto 120.80   13.63 lineto
122.73   18.24 lineto closepath
fill stroke}%
\special{"
    1.00    0.00    0.00 setrgbcolor
    0.40 setlinewidth
newpath  104.85  148.78 moveto   71.88   35.13 lineto stroke}%
\special{"
    1.00    0.00    0.00 setrgbcolor
0.0 setlinewidth newpath  88.36   91.95 moveto  84.05   86.05 lineto
 88.85   84.66 lineto closepath
fill stroke}%
\special{"
    1.00    0.00    0.00 setrgbcolor
    0.40 setlinewidth
newpath   71.88   35.13 moveto  160.00    0.00 lineto stroke}%
\special{"
    1.00    0.00    0.00 setrgbcolor
0.0 setlinewidth newpath 115.94   17.56 moveto 121.39   12.70 lineto
123.24   17.34 lineto closepath
fill stroke}%
\special{"
    1.00    0.00    0.00 setrgbcolor
0.0 setlinewidth
newpath   67.85   16.17 moveto   66.35   16.17 1.5 0 360 arc closepath
fill stroke}%
\special{"
    1.50 setlinewidth
newpath   94.18  151.46 moveto  105.82  148.54 lineto stroke}%
\put(0,6){\makebox(0,0)[bc]{\bm{O}}}
\put(165,0){\makebox(0,0)[lc]{\bm{s}}}
\put(66.35,11.17){\makebox(0,0)[tc]{\bm{i}}}
\put(111.49,143.58){\makebox(0,0)[lc]{\bf eye}}
\end{picture}

%% file: figures/im4.tex
\begin{picture}(170,220)(-10,-10)
\special{"
0.0 setlinewidth
newpath 2.5 0 moveto 0 0 2.5 0 360 arc closepath
gsave 0.0 setgray fill grestore stroke}%
\special{"
1.0 setlinewidth
newpath 78.78  -13.89 moveto 0 0 80 -10 100 arc stroke}%
\special{"
    1.00    0.00    0.00 setrgbcolor
0.0 setlinewidth
newpath  162.50    0.00 moveto  160.00    0.00 2.5 0 360 arc closepath
fill stroke}%
\special{"
    1.00 setlinewidth
newpath   47.87  164.85 moveto   42.88  144.29 lineto stroke}%
\special{"
    1.00 setlinewidth
newpath   47.87  164.85 moveto   58.44  146.52 lineto stroke}%
\special{"
    1.00    0.00    0.00 setrgbcolor
    0.40 setlinewidth
newpath   50.00  150.00 moveto   64.76   46.97 lineto stroke}%
\special{"
    1.00    0.00    0.00 setrgbcolor
0.0 setlinewidth newpath  57.38   98.48 moveto  55.88   91.33 lineto
 60.83   92.04 lineto closepath
fill stroke}%
\special{"
    1.00    0.00    0.00 setrgbcolor
    0.40 setlinewidth
newpath   64.76   46.97 moveto  160.00    0.00 lineto stroke}%
\special{"
    1.00    0.00    0.00 setrgbcolor
0.0 setlinewidth newpath 112.38   23.48 moveto 117.44   18.20 lineto
119.65   22.69 lineto closepath
fill stroke}%
\special{"
    1.00    0.00    0.00 setrgbcolor
    0.40 setlinewidth
newpath   45.05  149.29 moveto   63.91   48.12 lineto stroke}%
\special{"
    1.00    0.00    0.00 setrgbcolor
0.0 setlinewidth newpath  54.48   98.71 moveto  53.28   91.50 lineto
 58.20   92.41 lineto closepath
fill stroke}%
\special{"
    1.00    0.00    0.00 setrgbcolor
    0.40 setlinewidth
newpath   63.91   48.12 moveto  160.00    0.00 lineto stroke}%
\special{"
    1.00    0.00    0.00 setrgbcolor
0.0 setlinewidth newpath 111.95   24.06 moveto 116.98   18.75 lineto
119.22   23.22 lineto closepath
fill stroke}%
\special{"
    1.00    0.00    0.00 setrgbcolor
    0.40 setlinewidth
newpath   54.95  150.71 moveto   65.57   45.83 lineto stroke}%
\special{"
    1.00    0.00    0.00 setrgbcolor
0.0 setlinewidth newpath  60.26   98.27 moveto  58.47   91.18 lineto
 63.44   91.69 lineto closepath
fill stroke}%
\special{"
    1.00    0.00    0.00 setrgbcolor
    0.40 setlinewidth
newpath   65.57   45.83 moveto  160.00    0.00 lineto stroke}%
\special{"
    1.00    0.00    0.00 setrgbcolor
0.0 setlinewidth newpath 112.79   22.91 moveto 117.87   17.67 lineto
120.06   22.16 lineto closepath
fill stroke}%
\special{"
    1.00    0.00    0.00 setrgbcolor
0.0 setlinewidth
newpath   68.51   31.26 moveto   67.01   31.26 1.5 0 360 arc closepath
fill stroke}%
\special{"
    1.50 setlinewidth
newpath   44.06  149.15 moveto   55.94  150.85 lineto stroke}%
\put(0,6){\makebox(0,0)[bc]{\bm{O}}}
\put(165,0){\makebox(0,0)[lc]{\bm{s}}}
\put(67.01,26.26){\makebox(0,0)[tc]{\bm{i}}}
\put(63.44,146.52){\makebox(0,0)[lc]{\bf eye}}
\end{picture}

%% file: figures/bil1.tex
\begin{picture}(220,220)(-110,-110)
\special{"
0 1 0 setrgbcolor
0.0 setlinewidth
newpath   100.00 0 moveto 0 0   100.00 0 360 arc closepath
fill stroke}%
\special{"
0.7 0.3 0 setrgbcolor
    8.00 setlinewidth
newpath   104.00 0 moveto 0 0  104.00 0 360 arc
stroke}%
\special{"
    0.00    0.00    1.00 setrgbcolor
0.0 setlinewidth
newpath    98.84   24.93 moveto    96.84   24.93    2.00 0 360 arc closepath
fill stroke}%
\special{"
    0.20 setlinewidth
newpath    75.00    0.00 moveto    96.84   24.93 lineto stroke}%
\special{"
    0.20 setlinewidth
newpath     0.00   60.00 moveto    96.84   24.93 lineto stroke}%
\special{"
    0.00    0.00    0.00 setrgbcolor
0.0 setlinewidth newpath   85.92   12.47 moveto   83.28    5.65 lineto
  79.51    8.95 lineto closepath
fill stroke}%
\special{"
    0.00    0.00    0.00 setrgbcolor
0.0 setlinewidth newpath   48.42   42.47 moveto   55.73   42.48 lineto
  54.03   37.78 lineto closepath
fill stroke}%
\special{"
    0.00    0.00    1.00 setrgbcolor
0.0 setlinewidth
newpath   -72.64  -66.55 moveto   -74.64  -66.55    2.00 0 360 arc closepath
fill stroke}%
\special{"
    0.20 setlinewidth
newpath    75.00    0.00 moveto   -74.64  -66.55 lineto stroke}%
\special{"
    0.20 setlinewidth
newpath     0.00   60.00 moveto   -74.64  -66.55 lineto stroke}%
\special{"
    0.00    0.00    0.00 setrgbcolor
0.0 setlinewidth newpath    0.18  -33.28 moveto    5.44  -28.20 lineto
   7.47  -32.77 lineto closepath
fill stroke}%
\special{"
    0.00    0.00    0.00 setrgbcolor
0.0 setlinewidth newpath  -37.32   -3.28 moveto  -38.65  -10.46 lineto
 -42.96   -7.92 lineto closepath
fill stroke}%
\special{"
    1.00    1.00    1.00 setrgbcolor
0.0 setlinewidth
newpath    82.00    0.00 moveto    75.00    0.00    7.00 0 360 arc closepath
fill stroke}%
\special{"
    1.00    0.00    0.00 setrgbcolor
0.0 setlinewidth
newpath     7.00   60.00 moveto     0.00   60.00    7.00 0 360 arc closepath
fill stroke}%
\end{picture}

%% file: figures/bil2.tex
\begin{picture}(220,220)(-110,-110)
\special{"
0 1 0 setrgbcolor
0.0 setlinewidth
newpath   100.00 0 moveto 0 0   100.00 0 360 arc closepath
fill stroke}%
\special{"
0.7 0.3 0 setrgbcolor
    8.00 setlinewidth
newpath   104.00 0 moveto 0 0  104.00 0 360 arc
stroke}%
\special{"
    0.00    0.00    1.00 setrgbcolor
0.0 setlinewidth
newpath    72.71   70.71 moveto    70.71   70.71    2.00 0 360 arc closepath
fill stroke}%
\special{"
    0.20 setlinewidth
newpath    75.00    0.00 moveto    70.71   70.71 lineto stroke}%
\special{"
    0.20 setlinewidth
newpath     0.00   75.00 moveto    70.71   70.71 lineto stroke}%
\special{"
    0.00    0.00    0.00 setrgbcolor
0.0 setlinewidth newpath   72.86   35.36 moveto   75.77   28.65 lineto
  70.78   28.35 lineto closepath
fill stroke}%
\special{"
    0.00    0.00    0.00 setrgbcolor
0.0 setlinewidth newpath   35.36   72.86 moveto   42.36   74.93 lineto
  42.06   69.94 lineto closepath
fill stroke}%
\special{"
    0.00    0.00    1.00 setrgbcolor
0.0 setlinewidth
newpath    92.24   43.10 moveto    90.24   43.10    2.00 0 360 arc closepath
fill stroke}%
\special{"
    0.20 setlinewidth
newpath    75.00    0.00 moveto    90.24   43.10 lineto stroke}%
\special{"
    0.20 setlinewidth
newpath     0.00   75.00 moveto    90.24   43.10 lineto stroke}%
\special{"
    0.00    0.00    0.00 setrgbcolor
0.0 setlinewidth newpath   82.62   21.55 moveto   82.69   14.24 lineto
  77.97   15.91 lineto closepath
fill stroke}%
\special{"
    0.00    0.00    0.00 setrgbcolor
0.0 setlinewidth newpath   45.12   59.05 moveto   52.43   59.12 lineto
  50.76   54.40 lineto closepath
fill stroke}%
\special{"
    0.00    0.00    1.00 setrgbcolor
0.0 setlinewidth
newpath    45.10   90.24 moveto    43.10   90.24    2.00 0 360 arc closepath
fill stroke}%
\special{"
    0.20 setlinewidth
newpath    75.00    0.00 moveto    43.10   90.24 lineto stroke}%
\special{"
    0.20 setlinewidth
newpath     0.00   75.00 moveto    43.10   90.24 lineto stroke}%
\special{"
    0.00    0.00    0.00 setrgbcolor
0.0 setlinewidth newpath   59.05   45.12 moveto   63.69   39.48 lineto
  58.98   37.81 lineto closepath
fill stroke}%
\special{"
    0.00    0.00    0.00 setrgbcolor
0.0 setlinewidth newpath   21.55   82.62 moveto   27.19   87.27 lineto
  28.86   82.55 lineto closepath
fill stroke}%
\special{"
    0.00    0.00    1.00 setrgbcolor
0.0 setlinewidth
newpath   -68.71  -70.71 moveto   -70.71  -70.71    2.00 0 360 arc closepath
fill stroke}%
\special{"
    0.20 setlinewidth
newpath    75.00    0.00 moveto   -70.71  -70.71 lineto stroke}%
\special{"
    0.20 setlinewidth
newpath     0.00   75.00 moveto   -70.71  -70.71 lineto stroke}%
\special{"
    0.00    0.00    0.00 setrgbcolor
0.0 setlinewidth newpath    2.14  -35.36 moveto    7.23  -30.11 lineto
   9.42  -34.61 lineto closepath
fill stroke}%
\special{"
    0.00    0.00    0.00 setrgbcolor
0.0 setlinewidth newpath  -35.36    2.14 moveto  -36.10   -5.13 lineto
 -40.60   -2.94 lineto closepath
fill stroke}%
\special{"
    1.00    1.00    1.00 setrgbcolor
0.0 setlinewidth
newpath    82.00    0.00 moveto    75.00    0.00    7.00 0 360 arc closepath
fill stroke}%
\special{"
    1.00    0.00    0.00 setrgbcolor
0.0 setlinewidth
newpath     7.00   75.00 moveto     0.00   75.00    7.00 0 360 arc closepath
fill stroke}%
\end{picture}